\documentstyle[epsf,aaspp4]{article}

\lefthead{}
\righthead{}
\begin{document}
\title{Primordial Lithium Abundance as a Stringent
Constraint on the Baryonic Content of the Universe}

\author{Takeru Ken Suzuki$^1$, Yuzuru Yoshii$^2$, \& Timothy C. Beers$^3$}
\altaffiltext{1}{Department of Astronomy, School of Science,
University of Tokyo, Bunkyo-ku, Tokyo, 113-0033 Japan; Theoretical
Astrophysics Division, National
Astronomical Observatory, Mitaka, Tokyo, 181-8588 Japan;
stakeru@th.nao.ac.jp}
\altaffiltext{2}{Institute of Astronomy, School of Science,
University of Tokyo, Mitaka, Tokyo, 181-8588 Japan;
Research Center for the Early Universe, Faculty
of Science, University of Tokyo, Bunkyo-ku, Tokyo, 113-0033 Japan}
\altaffiltext{3}{Department of Physics and Astronomy, Michigan State
University, East Lansing, MI 48824, USA}

\begin{abstract}

We have refined the estimate of the primordial level of $^7$Li abundance to an
accuracy better than 10\%, based on high-precision Li abundances for
metal-poor halo stars, and a recent model of post-BBN (Big Bang
Nucleosynthesis) chemical evolution that provides a quantitative explanation of
the detected gentle ascent of the Spite Plateau for stars with metallicities
[Fe/H]$ > -3$.  Our maximum likelihood analysis obtains an estimate for the
primordial Li abundance of $A({\rm Li})_p=2.07^{+0.16}_{-0.04}$, after taking
into account possible systematic errors in the estimation of Li abundances,
with the exception of a still-controversial issue regarding stellar depletion.
The inferred value of $\eta$ (the baryon-to-photon number-density ratio in the
universe) based on this estimate is more consistent with that derived from the
set of reported ``low He'' + ``high D'' from extragalactic sites than that
derived from reported ``high He'' + ``low D'' measurements.  Since, within
current models of stellar depletion processes, it is difficult to account for
the observed very small scatter of Li abundance in metal-poor stars, our
estimate of $A({\rm Li})_p$ should be taken as an independent constraint on the
baryonic mass density parameter in the universe, giving
$\Omega_b h^2=(0.64-1.4)\times 10^{-2}$ with $h=H_0/100$ km
s$^{-1}$Mpc$^{-1}$. 
\end{abstract}


\section{Introduction}

The absolute abundances of $^4$He, $^2$D, and $^7$Li synthesized in the first
three minutes following the hot Big Bang provide the key to a determination of
the universal baryon density via its relationship to the $\eta$ parameter.  In
order to refine observational estimates of the primordial levels of these light
elements, previous attempts have been made to measure $^4$He and $^2$D
abundances in extragalactic sites, where potential problems associated with
correction for the effects of post-BBN chemical evolution could be minimized or
avoided.  However, the ``low'' value of $^{4}$He/H reported for metal-poor
extragalactic HII regions by Pagel et al. (1992) and Olive, Skillman,
\& Steigman (1997) stands in contrast to the ``high'' value of $^{4}$He/H
reported by Izotov \& Thuan (1998).   The ``high'' value for $^2$D/H
reported for high-redshift intergalactic HI clouds by Songaila et al. (1994),
Carswell et al. (1994), and Rugers \& Hogan (1996) appears at odds with the
``low'' values for $^2$D/H from Tytler et al. (1996), Burles \& Tytler
(1998ab), and Burles et al. (1999).  Detection of a high $^2$D/H abundance
(Webb et al. 1997; Tytler et al. 1999) for a gas cloud at rather low redshift
($z\simeq$0.7) has made the problem more complicated, because this is opposite
to the expectation, based on one-zone models of chemical evolution, that the
amount of $^2$D is should decrease following BBN as the result of various
destruction processes.

Given that two different values for both the primordial levels of $^4$He and
$^2$D, leading to two distinguishable values of $\eta$, remain tenable at
present, there is a pressing need for accurate estimate of the primordial level
of $^7$Li, which provides an independent constraint on $\eta$.  Obtaining such
a constraint has proven difficult for two reasons.  First, the prediction of
the primordial $^{7}$Li abundance is a non-linear function of $\eta$, which
formally permits the assignment of two values for $\eta$ at each level of
measured primordial $^7$Li.  Second, $^7$Li is several orders of magnitude less
abundant than the other two light elements, so that high-precision
observations of surface Li abundances are not possible except for nearby stars,
which might have experienced the effects of chemical evolution in the Galaxy.

Despite these apparent difficulties, ever since the discovery of a roughly
constant value of Li ($^{6}$Li + $^{7}$Li) abundance in a small sample of
metal-poor dwarf stars by Spite \& Spite (1982, the so-called ``Spite
Plateau''), many groups (e.g., Ryan et al 1996; hereafter RBDT; Bonifacio
\& Molaro 1997) have attempted to better determine the appropriate
primordial level of this element (see Spite 2000 for a recent review).  The
recent high precision (and homogeneously analyzed) data of Ryan, Norris,
\& Beers (1999, hereafter RNB) showed that the Spite plateau is in fact
incredibly ``thin,'' with an intrinsic star-to-star scatter in derived
Li abundance $\sigma < 0.02$ dex.

RNB also claimed the existence of a statistically significant slope of $A$(Li)
{\it versus} [Fe/H] in the Spite Plateau at low metallicity (first detected by
Thorburn 1994), apparently due to the influence of early Galactic chemical
evolution.  Ryan et al. (2000, hereafter RBOFN) have shown that the observed
slope in the Spite plateau can be used to {\it empirically} constrain the total
expected contribution from Galactic Cosmic Rays (GCRs) and supernovae (SNe).
These authors showed that a simple one-zone model for chemical evolution
produces a slope which is of similar magnitude to that which is observed.
However, their evaluation is necessarily tied to the presently uncertain
relationship between O and Fe abundances in the early Galaxy.  Furthermore,
their evaluation is based on a model which ignores the expected stochastic
nature of early star formation which is likely to apply during the first 10$^7$
to 10$^9$ years of chemical evolution in the early Galaxy.

In this paper we employ the SN-induced chemical evolution model presented by
Tsujimoto, Shigeyama, \& Yoshii (1999; hereafter TSY) and Suzuki, Yoshii, \&
Kajino (1999; hereafter SYK), which has the great advantage of treating the
production of heavy and light elements {\it consistently} in the inhomogeneous
early Galaxy (\S 2).  We recover the observed slope of $A$(Li) {\it versus}
[Fe/H], and obtain an estimate of the primordial level of lithium (\S
3).  We discuss the impact of our results in the context of the standard
BBN model (\S 4). 

\section{The Model for Galactic Chemical Evolution}

Recent observations of the most metal-deficient stars in the Galaxy reveal that
the elemental abundance patterns in these stars seem to reflect the
contributions from {\it single} SN events (Audouze \& Silk 1995; McWilliam et
al. 1995; Ryan, Norris, \& Beers 1996; Norris, Beers, \& Ryan 2000).  These
stars may have been formed in individual SN remnant (SNR) shells, at a time
when the interstellar gas was not well-mixed throughout the halo.  Based on
this scenario, TSY presented a SN-induced chemical evolution model which
successfully explains the observed large scatter of Eu abundances in very
metal-poor stars.  An important prediction of this model is that metallicity,
especially [Fe/H], cannot be used as an age indicator at early epochs.  Similar
models of early Galactic chemical evolution have been studied by Argast et al.
(2000).  The clear implication is that when one considers elemental abundances
of metal-poor stars a {\it distribution} of stellar abundances must be
constructed in the context of an explicit model, rather than taking the
evolution of elements assuming a well-mixed ISM, as is often done in simple
one-zone models (e.g., Fig.4 in TSY).

SYK extended the SN-induced chemical evolution model for analysis of the
evolution of the light elements produced by both primary and secondary
processes involving GCRs, and demonstrated that this model also reproduces the
observed trends of $^9$Be and B ($^{10}$B + $^{11}$B) data.  Thus, a SN-induced
chemical evolution model appears suitable for a self-consistent investigation
of the evolution of various elements in an inhomogeneous Galactic halo.

For the purpose of comparison with previous work, we here consider the
evolution of Li ($^{6}$Li + $^{7}$Li) as predicted using the same model
presented in SYK and TSY.  Spallative and fusion reactions of GCRs produce both
$^{6}$Li and $^{7}$Li, while the $\nu$-process of SNe produces $^{7}$Li alone.
The predictions of Woosley \& Weaver (1995) for progenitors of different mass
are used for the yields of the $\nu$-process, but the absolute values of these
yields are reduced in order to match the observed $^{11}$B/$^{10}$B ratio
(Vangioni-Flam et al. 1996; Vangioni-Flam et al. 1998).  The transport of GCRs
is calculated by the leaky box model (Meneguzzi et al. 1971), and the source
spectrum of energetic particles of each element associated with SNe is taken
from SYK:
\begin{displaymath}
\hspace{-1cm}\displaystyle
q_i(E,t){\propto}\frac{E+E_0}{[E(E+2E_0)]^{\frac{\gamma+1}{2}}}
\int_{\max ({m_t{\rm ,}\,
m_{SN,l}})}^{m_{u}}\hspace{-1.2cm}dm
\end{displaymath}
\begin{equation}
\label{eqn:crfx}
\times\{M_{Z_i}(m)+Z_{i,{\rm g}}(t)f_{\rm cr}M_{\rm sw}(m,t)\}
{\phi(m) \over A_im}{\dot M}_\ast ({t-\tau (m)}),
\end{equation}
where $E_0=931{\rm MeV}$, $M_{Z_i}(m)$ is the mass of the $i$-th heavy
element synthesized and ejected from a star with mass $m$, and
${\dot M}_\ast (t)$ is the star formation rate at time $t$.  GCRs are
assumed to come from both swept-up material (accelerated in the SN shock
front) and SN-ejecta; $f_{\rm cr}$ is a free parameter which represents
the contributions to GCRs from these two sources.  In order to reproduce
the observed data of $^9$Be (Boesgaard et al. 1999) and $^{6}$Li (Smith,
Lambert, \& Nissen 1998), the value of the GCR spectral index is set to
$\gamma=2.7$, and the GCR composition parameter is set to $f_{\rm cr}=
0.007$ (Suzuki \& Yoshii 2000).  This parameter corresponds to the
situation where 3.5\% of the total GCRs arise directly from SNe ejecta.
The absolute value of $q_i(E,t)$ is chosen so that the predicted $^9$Be
abundance agrees with log(Be/H)=$-13.5$ at [Fe/H]$=-3.0$ as observed by
Boesgaard et al. (1999).

\section{Results}

Figure 1 presents a comparison of the observations of $A$(Li) from the
compilation of RBDT, and the more recent observations of RNB, to the
theoretical prediction of the frequency distribution of long-lived stars in the
$A$(Li)--[Fe/H] plane. In this comparison we have used our best estimate of the
primordial Li abundance $A({\rm Li})_p=2.09$ (justified below), and have
selected only the data for stars with ${\rm T_{eff}>6000K}$ and [Fe/H]$ <
-1.5$.  Because the two data sets exhibit rather different random errors, we
plot the data of RNB only in the top panel of Fig. 1, and the data of both RNB
and RBDT in the bottom panel, with the same theoretical predictions overlayed.

The effects due to early Galactic chemical evolution of Li (which produces
levels in excess of $A({\rm Li})_p$) are seen already at [Fe/H]$ \sim -3$; the
predicted trend is quite consistent with the data.  Our model confirms the
existence of the slope of Li abundances toward the metal-rich side of the Spite
Plateau, a result which be tested further once a larger sample of precision
data are obtained for more metal-rich stars (Ryan et al. 2000).  It should be
emphasized that this increasing trend of Li abundance is naturally derived by
use of the same model which fits the data of the other light elements such as
$^6$Li, $^9$Be and B ($^{10}$B+$^{11}$B) in metal-poor halo stars. The average
excess of total Li over the primordial $^{7}$Li value amounts to 0.03 dex at
[Fe/H] $=-2.5$, 0.09 dex at [Fe/H] $=-2$, and 0.24 dex at [Fe/H] $=-1.5$.  We
note that the contribution of $^{7}$Li production through the $\nu$-process is
less than 10\% of the total production rate of Li, which indicates that most of
Li in our model of the early Galaxy is produced by GCR $\alpha + \alpha$
reactions.

Since our model gives a reasonable amount of Li produced by post--BBN 
processes, it can be used to predict the precise quantity of Li synthesized
during the BBN era.  We estimate the primordial Li abundance $A({\rm
Li})_p$ by constructing likelihood plots as a function of $A({\rm Li})_p$. To
establish this likelihood function, the data (including random errors) are
compared with the predicted stellar distribution for alternative values of
$A({\rm Li})_p$ in the $A$(Li)--[Fe/H] plane.  The predicted probability
distribution $P({\bf x})$ for stars located at ${\bf x}$=([Fe/H],
$A({\rm Li}))$ is then calculated independently on each [Fe/H] grid.
If we take $g({\bf x}-{\bf x}_i, \sigma_{{\bf x}_i})$ to represent
the Gaussian distribution at ${\bf x}$ for the $i$-th star observed at
${\bf x}_i$ with errors of $\sigma_{{\bf x}_i}$, the likelihood
$L(A({\rm Li})_p)$ on each $A({\rm Li})_p$ can be calculated as
\begin{equation}
\label{eq:lkl}
\log\{L(A({\rm Li})_p)\}
={\Sigma}_{i}
\log\{\int_{\rm all\; space} \hspace{-0.5cm} P({\bf x})
g({\bf x}-{\bf x}_i, \sigma_{{\bf x}_i}) d^2{\bf x} \} \;\; .
\end{equation}
The inset of the top panel of Fig. 1 shows the likelihood function established
by using the data in RNB and RBDT.  Based on our likelihood analysis, the 95\%
confidence region on the primordial Li abundance occurs for
\begin{equation}
\label{eq:prpr}
A({\rm Li})_p=2.09^{+0.01}_{-0.02} \; .
\end{equation}
We have examined the results using the RNB data and the RBDT data respectively,
but there is little ($ < 0.005$ dex), if any, difference between them in the
derived estimate of $A({\rm Li})_p$.

So far we have not included any {\it systematic} errors in the determination of
Li abundances which might arise from a variety of sources.  Following RBOFN, we
now explicitly include estimates of these errors, {\it except} for the
still-controversial issue of stellar depletion.  The estimated value and
confidence interval for $A({\rm Li})_p$ then becomes
\begin{equation}
\label{eq:cnspl}
A({\rm Li})_p=2.07^{+0.16}_{-0.04}\; ,
\end{equation}
where we have considered the errors arising from the use of 1-D model
atmospheres (+0.10 dex), the adopted convection treatment (+0.08 dex), non-LTE
effects ($-0.02\pm 0.01$ dex), and uncertainties in adopted $gf$-values ($\pm
0.04$ dex).  The estimation of stellar abundances of Fe and Li also depends
crucially on the adopted temperature scale.  However, the [Fe/H] values we have
taken from RNB and RBDT come from a variety of sources, with slightly
different temperature scales, so a global correction is not practical.  Note
that a 100 K difference in temperature scales yields a 0.065 dex difference in
derived lithium abundance (RNB).  In particular, if one adopts the temperature
scale of Alonso, Arribas, \& Martinez-Roger (1996), which is on average 120K
{\it hotter} than the scale in RNB, the derived Li abundances might be on the
order of +0.08 dex higher.  Except for the $-0.02$ dex offset from non-LTE
effects, we take quadratic sums for these positive and negative errors
separately, and have derived $A({\rm Li})_p$ in equation 4 as our estimate with
the 95\% confidence levels including systematic errors.

\section{Summary and Discussion}

The importance of our new estimates of the primordial level of $^7$Li is that
they are derived, for the first time, from a self-consistent model which
explains both the $^6$LiBeB observations and the small, but {\it real},
increasing trend of Li appearing at [Fe/H]$ > -3$.  Figure 2 shows our
preferred value of $A({\rm Li})_p=2.07^{+0.15}_{-0.04}$ with a horizontal line,
and the allowed 95\% confidence intervals by a box, along with previous
estimates of $^4$He and $^2$D.  The theoretical prediction of standard BBN
(Thomas et al. 1994; Fiorentini et al. 1998) is superposed.  It is interesting
to note that our preferred value lies at the very bottom of the valley of the
function of $^7$Li abundance against $\eta$, which means that we can assign a
single value for $\eta$, independent of the results from $^4$He and $^2$D.  Our
analysis indicates $\eta({\rm Li})=(1.7-3.9)\times 10^{-10}$, which corresponds
to a universal baryonic density parameter\footnote{If taking into accout
uncertainties ($1\sigma$ errors) of nuclear 
reaction rates for the theoretical BBN calculation, the constraints
become $\eta({\rm Li})=(1.4-4.5)\times 10^{-10}$ and $\Omega_b
h^2=(0.53-1.7)\times 10^{-2}$.} $\Omega_b 
h^2=(0.64-1.4)\times 10^{-2}$ with the Hubble constant expressed as
$h=H_0/100$ km s$^{-1}$Mpc$^{-1}$.

The range of $\eta({\rm Li})$ we obtain appears to agree best with that
inferred from the reported ``low $^4$He'' + ``high $^2$D'' measured from
extragalactic sites, rather than the pair of reported ``high $^4$He'' + ``low
$^2$D''.  This range is also consistent with that inferred from the primordial
value of $^4$He obtained from recent observations of HII regions in the
Magellanic Clouds (Peimbert \& Peimbert 2000).

However, the possible effects of stellar depletion, which are not taken into
account here, might still play a role.  If one adopts the reported ``high
$^4$He'' and ``low $^2$D'' values as the correct ones with which to estimate
$\eta$, then it follows that $^7$Li has been destroyed through stellar
evolution during the long lifetimes of metal-poor halo stars.  From our Li
constraint one might conclude that  Li has been depleted by a factor of
$2.7^{+0.7}_{-1.4}$ as a result of stellar processing.

The required depletion factor, $\sim$ 2$-$3, is much larger than the prediction
($ < 1.2$; Deliyannis et al. 1990) of so-called standard stellar evolution
models which only take into account classical surface convection as the origin
of mixing in stellar interiors.  In order to destroy lithium in deeper and
hotter regions extra mixing processes are necessary.  Among such processes,
stellar rotation is thought to be one of the most effective candidates, but
depletion factors $\sim 1$ dex inferred by rotation-induced mixing depend
sensitively on the initial conditions of stellar rotation (Pinsonnealt et al.
1992).  So, in general, this process predicts that one might expect to see a
scatter about the Spite Plateau as large as the depletion factor itself,
reflecting star-to-star differences in stellar rotation and other properties.
Although recent models of rotation-induced mixing obtain more moderate
depletion factors $\sim 0.2-0.4$ dex (Pinsonnealt et al. 1999), the expected
scatter in the Spite Plateau would still seriously contradict the very small
intrinsic scatter ($\sigma < 0.02$ dex) observed by RNB.

All the above considerations indicate that, unless a novel process which is
capable of significant {\it and uniform} depletion of stellar Li abundance
is identified,  our preferred value of $A({\rm Li})_p$ should be taken as a
stringent constraint on $\eta$ or $\Omega_b$.  The inferred value of the
baryonic contribution to the density parameter is $\Omega_b\sim 0.01-0.02$
($h=0.75$).

This work has been supported in part by the Grant-in-Aid for the
Center-of-Excellence research (07CE2002) of the Ministry of Education,
Science, Sports, and Culture of Japan.  TCB acknowledges partial
support for this work from grant AST 95-29454 from the National Science
Foundation.

\newpage

\begin{figure}[h]
\epsfxsize=9cm
\epsfysize=12cm
\epsfbox{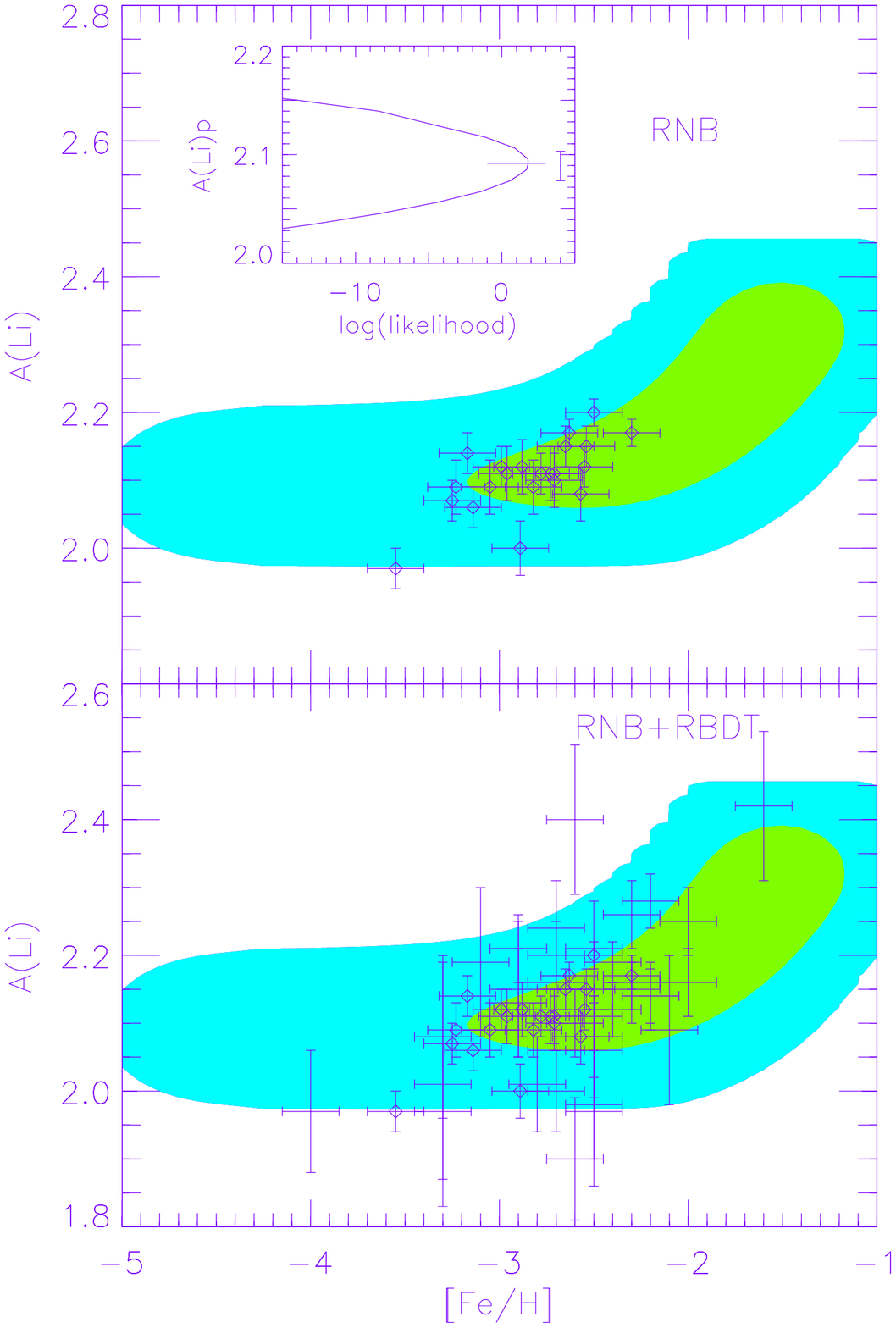}
\caption{Frequency distribution of long-lived stars in the $A({\rm
Li})$--[Fe/H] plane, convolved with a Gaussian having $\sigma = 0.03$ dex for
$A({\rm Li})$ and $\sigma = 0.15$ dex for [Fe/H].  The primordial lithium
abundance is chosen to be $A({\rm Li})_p=2.09$, which is the best fit value in
our model.  The two contour lines, from inside out, correspond to those of
constant probability density $10^{-4}$ and $10^{-9}$ in unit area of
$\Delta$[Fe/H]=0.1$\times\Delta$A(Li)=0.002. The inset shows the likelihood as
a function of $A({\rm Li})_p$.  The horizontal bar indicates the value of
$A({\rm Li})_p$ that gives the maximum likelihood, and the vertical bar shows
the range of 95\% confidence limits.  The crosses with and without lozenge
represent the data taken from RNB and RBDT, respectively, for stars with ${\rm
T_{eff}>6000K}$ and  [Fe/H]$<-1.5$.  The top panel shows a comparison of the
model with the data of RNB only, and the bottom panel for the combined data of
RNB and RBDT.}
\end{figure}

\newpage

\begin{figure}[h]
\epsfxsize=9cm
\epsfysize=12cm
\epsfbox{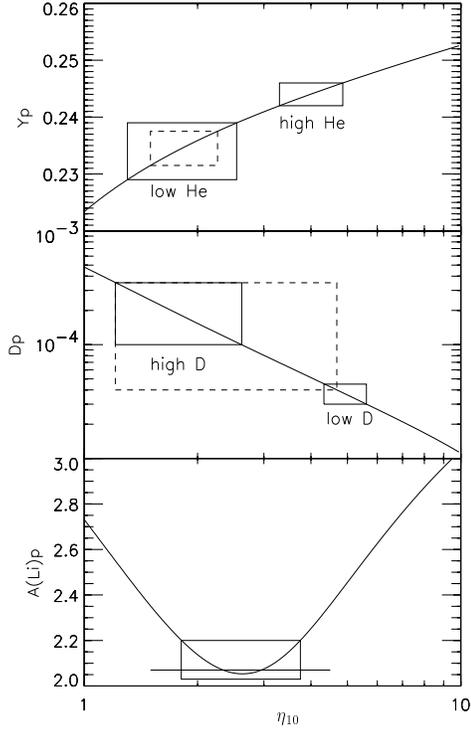}
\caption{Primordial abundances of $^4$He, $^2$D and $^7$Li as a function of
$\eta_{10} (\equiv\eta/10^{-10})$.  In the $A({\rm Li})-\eta$ diagram (bottom
panel), our determined $^7$Li value $A({\rm Li})_p=2.07^{+0.15}_{-0.04}$ is
shown by a horizontal line and the allowed region with errors by the box.  For
purposes of comparison, observational estimates of the primordial abundances of
$^4$He and $^2$D are also shown.  Their reference values of ``low $^4$He,''
``high $^4$He,'' ``low $^2$D,'' and ``high $^2$D'' are taken from Olive et al.
(1997), Izotov \& Thuan (1998), Burles \& Tytler (1998ab), and Rugers \& Hogan
(1996)$^5$, respectively.  The dashed box for ``low $^4$He''shows the result 
derived from observations of HII regions in the Magellanic Clouds (Peimbert \&
Peimbert 2000). The dashed box for ``high $^2$D'' shows the allowed region
based on a more conservative analysis of $^2$D for quasar absorption line
measurements (Songaila et al. 1997).}
\end{figure}
\altaffiltext{5}{The ``high $^2$D'' was obtained through the
observation of the cloud at $z=2.80$ in front of the quasar,
Q0014+813. However, Burles et al. (1999) pointed out that the observed
spectrum of this cloud could also be fitted with the model having
D/H$=$0.}

\end{document}